\documentclass[aps,prb,reprint,superscriptaddress,longbibliography]{revtex4-2}

\usepackage{graphicx}
\usepackage{amsmath,amssymb}
\usepackage{bm}
\usepackage{hyperref}
\usepackage{orcidlink}

\graphicspath{{figures/}}

\begin{document}

\title{Collective Phonon Mixing and Eigenvector Transport Under Isotope Substitution}

\author{Jeff Armstrong\,\orcidlink{0000-0002-8326-3097}}
\email{jeff.armstrong@stfc.ac.uk}
\affiliation{ISIS Neutron and Muon Source, STFC, Rutherford Appleton Laboratory, Harwell Campus, Didcot, OX11 0QX, UK}
\affiliation{Department of Chemistry, University of Bath, Bath, BA2 7AY, UK}

\author{Hamish Cavaye\,\orcidlink{0000-0002-3540-0253}}
\affiliation{ISIS Neutron and Muon Source, STFC, Rutherford Appleton Laboratory, Harwell Campus, Didcot, OX11 0QX, UK}

\author{Pankaj Sharma\,\orcidlink{0000-0003-2319-260X}}
\affiliation{United Kingdom Atomic Energy Authority, Culham Campus, Abingdon, Oxfordshire OX14 3DB, United Kingdom}

\author{Matthew E. Potter\,\orcidlink{0000-0001-9849-3306}}
\affiliation{Department of Chemistry, University of Bath, Bath, BA2 7AY, UK}

\begin{abstract}
Isotopic substitution modifies nuclear masses without altering the electronic potential energy surface to first order and is therefore often interpreted as a simple rescaling of vibrational frequencies. In solids with dense phonon manifolds, however, mass substitution acts as a parametric Hermitian deformation of the mass-weighted dynamical matrix, generating a continuous family of eigenproblems whose eigenvectors can undergo substantial rotation within coupled subspaces. Here we investigate protiated and deuterated ZIF-8 using inelastic neutron scattering and density functional theory lattice-dynamics calculations. While many vibrational modes exhibit near-ideal mass scaling and preserve their character across isotopic endpoints, modes embedded in spectrally congested regions display pronounced redistribution of vibrational character that cannot be inferred from frequency shifts alone. Because inelastic neutron scattering intensity is directly weighted by hydrogen displacement amplitude, spectral sparsity and congestion provide experimental indicators of predictable frequency renormalisation or susceptibility to qualitative eigenvector reorganisation under deuteration. To establish physically meaningful mode correspondence, we develop an adiabatic eigenvector-continuation framework with overlap-based tracking and explicit stability diagnostics. These results show that vibrational identity in complex framework materials is best understood as a continuous trajectory in eigenvector space and provide a general framework for analysing isotope-induced spectral flow in dense phonon systems.
\end{abstract}
\maketitle
\section{Introduction}
Isotopic substitution is widely used in neutron scattering because it modifies nuclear masses without, to first order, altering the electronic potential energy surface. This is exploited from a structural perspective in a range of ways. Due to the strongly modified coherent scattering cross-section ($\sigma_{Hcoh}\sim1.8$~barn $\to$ $\sigma_{Dcoh}\sim5.6$~barn), deuteration enables high sensitivity diffraction studies of organic crystals\cite{Bull2019_C9CE01028A}. Additionally, small angled scattering and reflectometry studies use these differences in coherent scattering as a contrasting method to discern nanoscopic structural boundaries\cite{Wu2025_SANS_nanoparticle_interfaces, Aoki2021_Langmuir_37_196}. In contrast, deuteration from a neutron spectroscopy perspective acts to drastically reduce the incoherent scattering cross section ($\sigma_{Hinc}\sim82.0$~barn $\to$ $\sigma_{Dinc}\sim7.6$~barn). Here again deuteration is often used as a contrasting species for peak identification of individual molecular groups\cite{Colognesi2025_ChemPhys597_112773}, and it may also be used as a means for tracking the temporal evolution of phonon bands during isotopic reactions\cite{Armstrong2021_VibrationalMotions}. More accurately however, this does not simply provide a contrasting mask, as it also acts to perturb the vibrational frequencies of the spectra. In its simplest form, replacing protium ($^1$H) with deuterium ($^2$H) leads to the familiar mass-scaling relation $\omega \propto 1/\sqrt{m}$ for an isolated harmonic oscillator. While this intuition is adequate for weakly coupled or spectrally sparse systems, it becomes incomplete in extended solids and molecular materials where vibrational states form dense, interacting manifolds. In such systems, the central question is not simply how far a peak shifts in frequency, but how vibrational identity evolves when the mass matrix is perturbed.

Within harmonic lattice dynamics, isotope substitution enters exclusively through the diagonal mass matrix ($\boldsymbol{M}$) and therefore acts as a deformation of the mass-weighted dynamical matrix,
\begin{equation}
\mathbf{D} = \mathbf{M}^{-1/2}\boldsymbol{\Phi}\mathbf{M}^{-1/2},
\end{equation}
where $\boldsymbol{\Phi}$ is the force-constant matrix. Changing the masses while holding $\boldsymbol{\Phi}$ fixed generates a continuous family of Hermitian eigenproblems. Although the eigenvalues evolve smoothly under such a perturbation, the associated eigenvectors need not preserve a simple one-to-one correspondence between isotopic endpoints. In spectrally congested regions, even modest mass-induced perturbations can drive substantial rotation of eigenvectors within coupled subspaces, leading to avoided crossings and redistribution of vibrational character. As a consequence, continuity of frequency ordering does not imply continuity of vibrational identity.

This distinction is not merely formal. Metal--organic frameworks (MOFs), with their large unit cells and mixed organic--inorganic composition, exhibit high densities of vibrational states across broad energy ranges. Framework modes influence pore geometries, adsorption-site configurations\cite{Mor2025Langmuir,Bahadur2025JPCL}, and the entropic components of gas uptake\cite{Butler2019_ChemMater31_8366}, all of which are sensitive to the detailed displacement patterns of the underlying phonons. More generally, in molecular solids and organic semiconductors, electron--phonon coupling depends not only on phonon frequencies but on the spatial structure of the vibrational eigenvectors that modulate transfer integrals and site energies. In both contexts, understanding how vibrational character reorganises under mass perturbation is therefore of direct physical relevance.
Inelastic neutron scattering (INS) is particularly well suited to this task.
Unlike optical spectroscopies, INS probes vibrational eigenstates without dipole or Raman selection rules and is strongly weighted toward hydrogen motion, through its large incoherent scattering cross section. 
Agreement between neutron-weighted density functional theory lattice-dynamics (DFT-LD) calculations and INS therefore provides a stringent test of both vibrational frequencies and displacement patterns\cite{Filippov2019_JPCC123_28631}, and the associated thermodynamic free energies\cite{Kieslich2018_ChemMater30_8782}.

Crucially, because one-phonon INS intensity is proportional to hydrogen displacement amplitude (modulated by the Debye–Waller factor), peak intensities in the protiated spectrum encode quantitative information about the projection of vibrational eigenvectors onto the perturbed mass subspace.
As the leading-order isotope-induced frequency shift depends on this same projection, spectrally isolated peaks with large hydrogen-weighted intensity can be expected to exhibit correspondingly large and predictable shifts under deuteration.
Conversely, in spectrally congested regions where multiple neighbouring modes share substantial hydrogen displacement weight, isotope substitution may promote collective redistribution of vibrational character. INS therefore provides not only a measure of vibrational frequencies, but an experimental diagnostic of the expected stability or reorganisation of displacement patterns under controlled mass perturbation.

In the present work we investigate protiated and deuterated ZIF-8 using INS measurements combined with DFT-LD calculations. To establish physically meaningful mode correspondence between isotopic endpoints, we treat isotope substitution as a continuous mass deformation and develop an adiabatic eigenvector-continuation framework. By propagating the eigenvectors of the mass-weighted dynamical matrix along this parameter and enforcing correspondence via overlap-based tracking with explicit stability diagnostics, we resolve ambiguities that arise in spectrally congested regions. This approach reveals where vibrational identity is preserved, where it is redistributed across coupled manifolds, and how interpolation resolution governs the apparent behaviour of near-degenerate modes. The analysis provides a general eigenvector-resolved framework for understanding isotope-induced spectral flow in dense phonon systems.
\section{Methods}
\subsection{Inelastic Neutron Spectroscopy Measurements}
Powder samples of the metal–organic framework ZIF-8 were measured in fully protiated and fully deuterated forms. The protiated and deuterated specimen masses were 2.325~g and 1.517~g respectively.
 INS measurements were performed on the TOSCA spectrometer at the ISIS Neutron and Muon Source, Rutherford Appleton Laboratory, UK\cite{Pinna2018TOSCA,TOSCA_ISIS}. TOSCA is an indirect-geometry neutron spectrometer optimized for measurements of vibrational dynamics in organic molecular crystals, providing high resolution across a broad energy-transfer range. All measurements were conducted at a temperature of 10~K. Low-temperature conditions minimize the Debye-Waller effect, thermal population factors, and anharmonic broadening, thereby enabling a direct comparison with DFT-LD calculations. Owing to the large difference in neutron scattering cross-sections between protiated and deuterated samples, the measurement times were approximately 3~h and 14~h respectively, to achieve comparable signal-to-background ratios.

Time-of-flight data was reduced within the Mantid framework\cite{Arnold2014NIM}, and empty cell backgrounds were subtracted from both data sets. The resulting spectra are presented in units of energy transfer and directly compared with neutron-weighted DFT-LD calculations.
\subsection{Density functional theory calculations}
DFT calculations were performed to obtain the equilibrium structure and
vibrational properties of pristine ZIF-8. All calculations were carried out using the Vienna \emph{ab
initio} Simulation Package (VASP)\cite{Kresse1996PRB,Kresse1996CMS}. The exchange--correlation energy
was treated within the generalized gradient approximation using the Perdew--Burke--Ernzerhof (PBE)
functional\cite{Perdew1996PRL}. Core--valence interactions were described using the
projector-augmented-wave (PAW) method,\cite{Kresse1999PRB} and a plane-wave basis set with an energy cutoff of 550~eV was employed throughout. Long-range dispersion interactions were included using
Grimme’s D3 correction with Becke--Johnson damping (PBE-D3)\cite{Grimme2011JCC}.

Structural optimization was performed on a pristine ZIF-8 unit cell Zn$_{12}$(C$_{4}$H$_{5}$N$_{2}$)$_{24}$, with both atomic positions and lattice parameters fully relaxed.
Phonon calculations were carried out for the optimized structure using density functional perturbation
theory (DFPT). From the calculated phonon eigenvalues and eigenvectors, INS spectra were simulated using
the AbINS package\cite{Dymkowski2018PhysB} and further processed using the Mantid
framework\cite{Arnold2014NIM}. Vibrational eigenmodes and atomic displacements were visualized using
Jmol\cite{Hanson2010JAC}. Combinations and overtones are presented in the direct comparison with INS data, but only fundamental modes are presented in the rest of the manuscript to ease unique mode assignment.
\section{Results}
\subsection{Experimental validation of Lattice Dynamics and direct observations}
\begin{figure}[!h]
 \centering
 \includegraphics[width=0.95\linewidth]{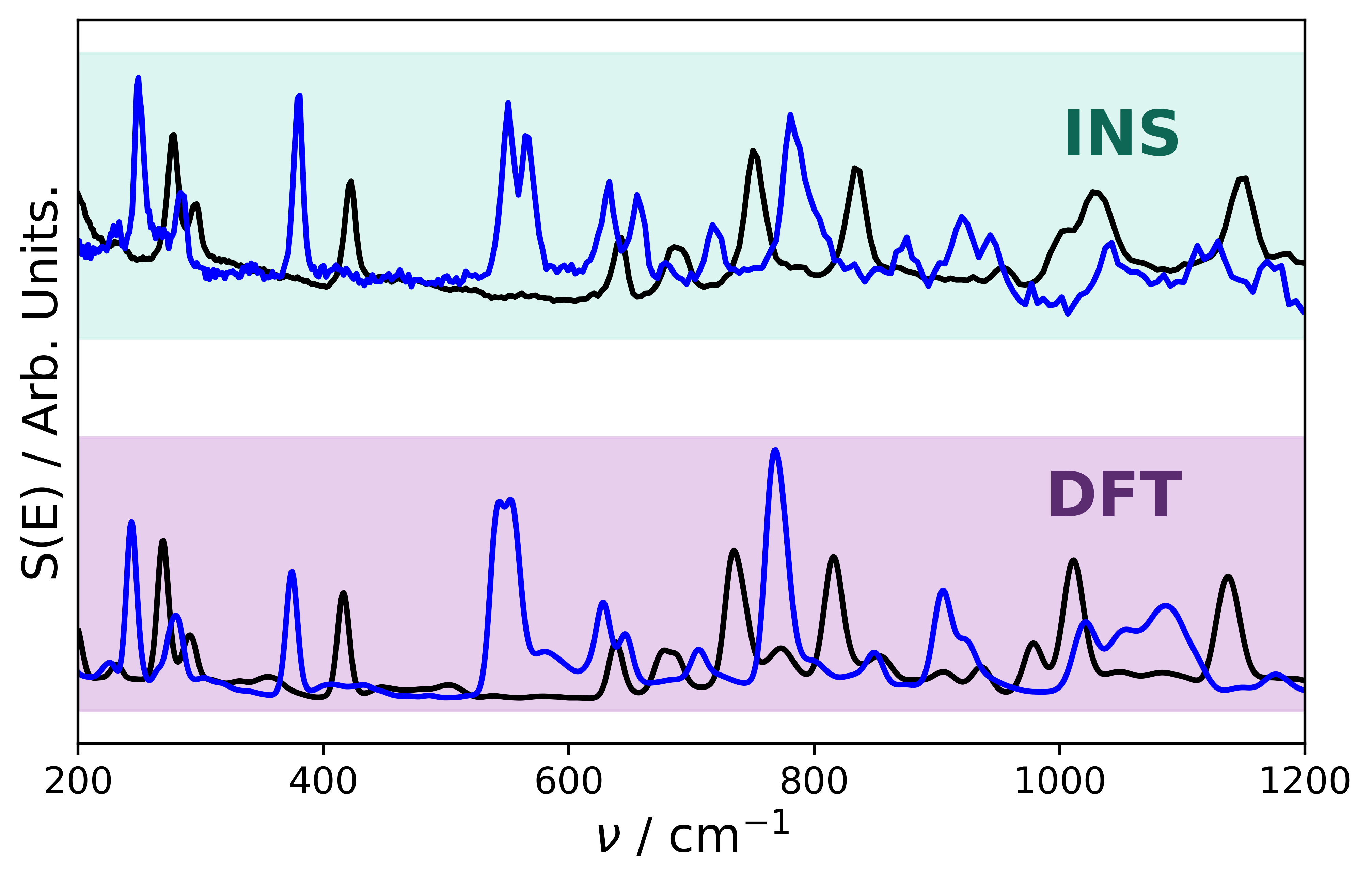}
 \caption{Experimental INS spectra (top) and neutron-weighted calculated spectra (bottom). Protiated ZIF-8 is shown in black and deuterated ZIF-8 in blue.}
 \label{exptvstheory}
\end{figure}

The experimental INS spectra of hydrogenated and deuterated ZIF-8 are shown in Fig.~\ref{exptvstheory}, together with neutron-weighted DFT-LD. The calculations reproduce peak positions and relative intensities across the vibrational fingerprint region, confirming that the DFT-LD force constants accurately describe the vibrational eigen-problem relevant to neutron scattering.

The isotope-induced spectral response is strongly energy dependent. Between approximately 200 and 450~cm$^{-1}$, several modes exhibit near-ideal mass renormalization (peak positions shift downward while spectral shape and intensity are largely preserved). A representative example occurs for the protiated frequency at $\sim$416~cm$^{-1}$, which migrates to $\sim$374~cm$^{-1}$ for the deuterated system.

In contrast, the region 450–1200~cm$^{-1}$ displays predominantly non-ideal behaviour. Peak shifts are non-uniform and frequency matching becomes ambiguous. Because this fingerprint region contains a high density of vibrational states, it provides the most stringent test of vibrational identity under mass perturbation.

\subsection{Direct comparison of H and D eigenvectors}
We now proceed to identify displacement pattern changes between the protiated and deuterated endpoints. Normal modes are obtained from the mass-weighted eigenproblem
\begin{equation}
\mathbf{D}\mathbf{q}_i = \omega_i^2 \mathbf{q}_i,
\label{eq:eig}
\end{equation}

where the eigenvectors $\mathbf{q}_i$ are dimensionless mass-weighted normal-mode vectors, normalized such that $\mathbf{q}_i^{\mathrm T}\mathbf{q}_j = \delta_{ij}$.
To quantify comparison between H and D isotopes, we compute the mass-weighted overlap between their respective eigenvectors,
\begin{equation}
O_{ij}
=
\left|
\left(\mathbf{q}_i^{(\mathrm H)}\right)^{\mathrm T}
\mathbf{q}_j^{(\mathrm D)}
\right|.
\label{eq:overlap}
\end{equation}

\begin{figure}[!h]
 \centering
 \includegraphics[width=0.95\linewidth]{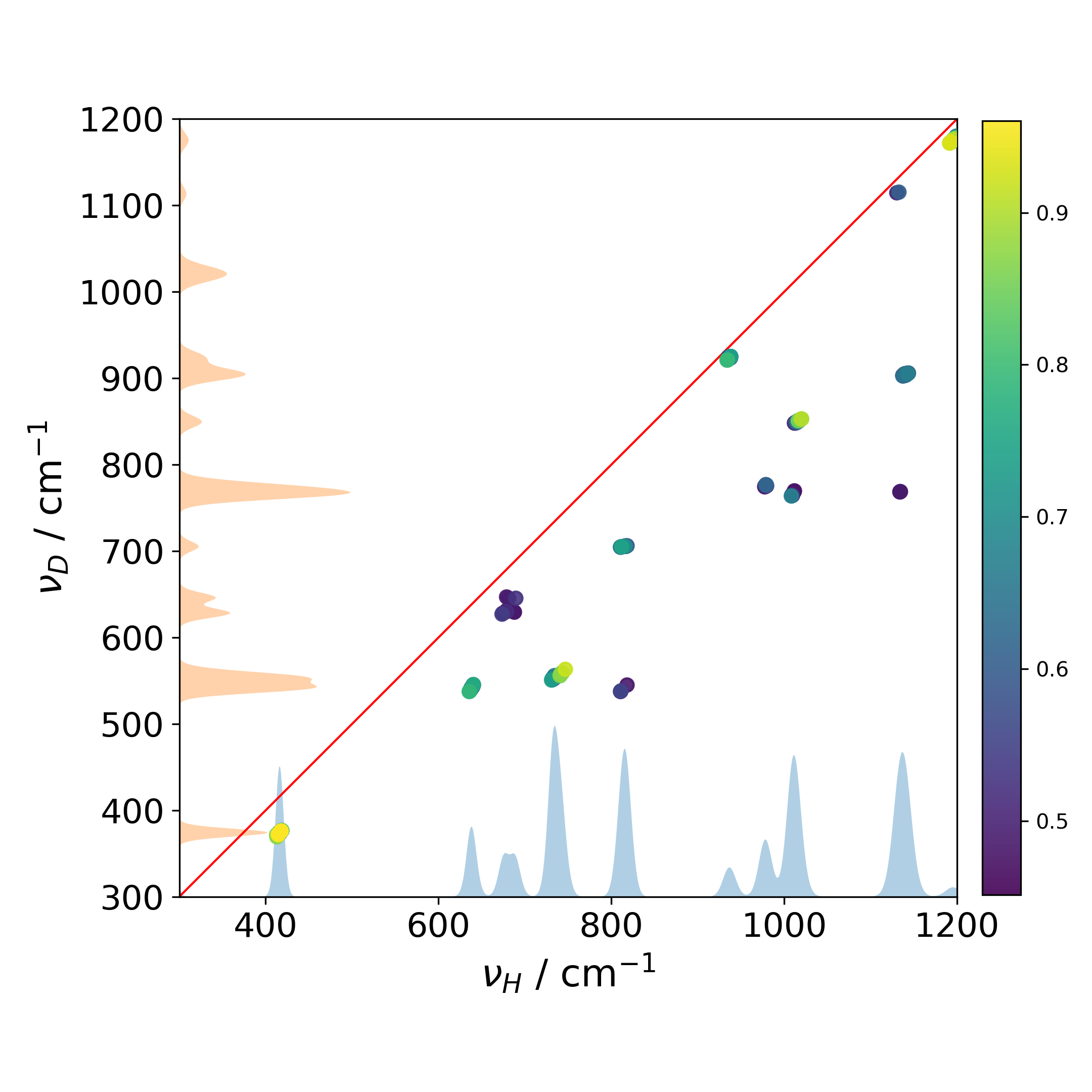}
 \caption{Mass-weighted overlap map for modes between 300 and 1200~cm$^{-1}$. $O_{ij}$ is represented by the color axis of the plot. Only pairs with $O_{ij} > 0.45$ are shown. Fundamental mode neutron projected DFT spectra for H and D are shown on the x and y axis respectively to aid visual band assignment.}
 \label{vector_overlaps}
\end{figure}

The overlap map for modes in the 300–1200~cm$^{-1}$ range is shown in Fig.~\ref{vector_overlaps}. In spectrally sparse regions, each hydrogenated mode exhibits a single dominant overlap with a unique deuterated mode (e.g., 416~cm$^{-1}$), forming an approximately one-to-one mapping across endpoints. In contrast, within the congested fingerprint region, many hydrogenated modes exhibit significant overlap with multiple deuterated modes, indicating redistribution of vibrational character and leading to ambiguity in assignments based solely on endpoint mass-weighted displacement patterns. Moreover, within a given congested band, the overlap metric itself exhibits substantial variation even across modes of similar frequency. This broad distribution of overlap values reflects collective mixing within near-degenerate manifolds and demonstrates that the magnitude of the endpoint overlap alone is not a reliable diagnostic of vibrational identity in dense spectral regions.

\begin{figure}[!h]
 \centering
 \includegraphics[width=0.95\linewidth]{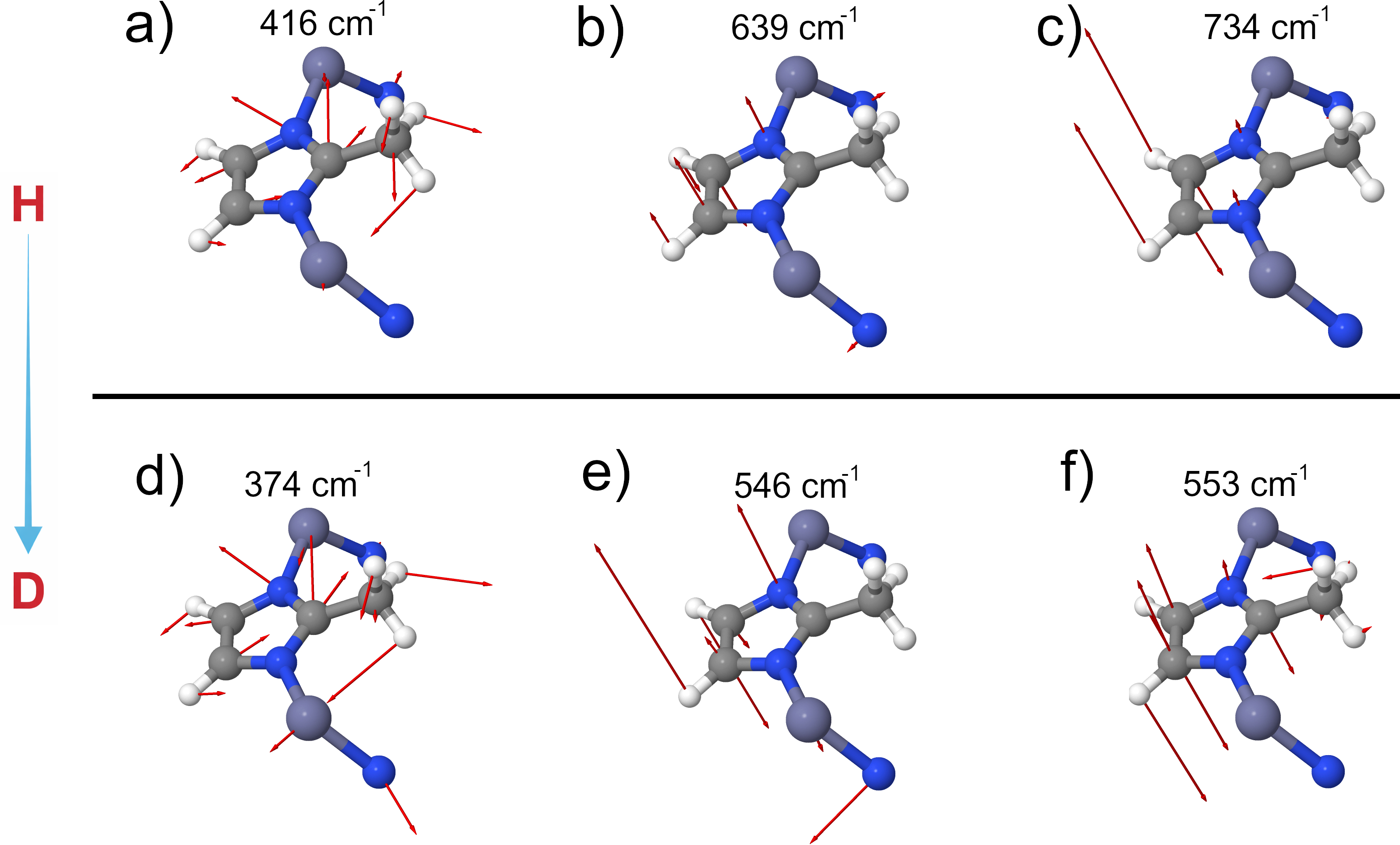}
 \caption{Mass-weighted displacement patterns for representative modes. Panels a), b) \& c) show protiated ZIF-8 modes at 416, 639, and 734~cm$^{-1}$, respectively. Panels d), e) \& f) show the corresponding deuterated modes at 374, 546, and 553~cm$^{-1}$.}
 \label{snaps}
\end{figure}

To gain some qualitative insight into these behaviors, we examine representative displacement patterns(Fig.~\ref{snaps}) of H and D bands which broadly show a sizable overlap metric. All displacement fields are shown in mass-weighted coordinates, $\mathbf{q}=\mathbf{M}^{1/2}\mathbf{u}$ ($\mathbf{u}$ representing the unweighted displacement), allowing direct comparison between protiated and deuterated systems without trivial amplitude rescaling due to mass changes. For ease of visualisation, we present a single molecular linker and two connected nodes from the chosen eigenvector. Note that modes cluster into distinct bands, and within these bands the range of overlaps was seen to vary substantially between protiated and deuterated variants, thus the illustrations to be viewed as representative examples of a given band.

Panels a) and d) compare the protiated 416~cm$^{-1}$ mode with its deuterated counterpart at 374~cm$^{-1}$. The spatial distribution of atomic motion, phase relationships, and participating atoms are very similar. Only the overall frequency is reduced. The mass-weighted displacement pattern is therefore well preserved, indicating that vibrational identity is maintained under isotopic substitution, which is in line with it residing in comparatively sparse spectral region.

Panels b) and e) compare a protiated mode at 639~cm$^{-1}$ mode with a deuterated mode at 546~cm$^{-1}$. Here the same atomic groups remain involved in the motion; however, the mass-weighted displacement amplitudes are redistributed significantly. Relative amplitudes among hydrogen, carbon, and nitrogen atoms change, and the balance of motion across the framework is altered. Although no entirely new atomic species become active, the eigenvector has rotated within its local subspace, leading to substantial quantitative reorganisation of vibrational character.

Panels c) and f) illustrate an even stronger deviation from ideal behaviour. The protiated 734~cm$^{-1}$ mode transforms into a deuterated mode at 553~cm$^{-1}$ with both amplitude redistribution and qualitative change in motion. In addition to altered relative displacements, new atomic participation becomes evident: a methyl rocking component appears in the deuterated case that is not prominent in the hydrogenated mode. This represents genuine subspace reorganisation rather than simple amplitude modulation.

These three examples therefore span a progression from near-ideal behaviour (A–D), to quantitative eigenvector redistribution (B–E), to qualitative mixing involving new atomic motion (C–F). More broadly this highlights the inherent ambiguity of such an end point analysis, as there is no reason apriori to assume that a deuterated endpont will share qualitative eigenvector character with a protiated counterpart. A deeper analysis of the eigenvector transport is required in order to understand the true correspondence of modes between these endpoints.

\subsection{Perturbative interpretation of frequency shifts, mixing, and spectral intensity}
To understand the transport between these isotopic endpoints, we employ a first-order perturbation theory approach to the mass deformation (See appendix\ref{sec:mainderive} for detailed derivation). The transport of frequency can be understood through the leading-order change in squared frequency, given by the diagonal matrix element in the eigenbasis,
\begin{equation}
\delta \omega_i^2
=
\mathbf{q}_i^{\mathrm T}\,\delta\mathbf{D}\,\mathbf{q}_i.
\label{eq:domega}
\end{equation}
Because $\delta\mathbf{D}$ arises from modification of the mass matrix, this term depends explicitly on the projection of the eigenvector onto the perturbed atomic coordinates. 
Modes with large displacement amplitude on hydrogen sites therefore exhibit the largest isotope-induced frequency shifts.
In the hydrogen dominated incoherent regime, the one-phonon intensity for mode $i$ follows the relation
\begin{equation}
I_i(Q,\omega)
\propto
e^{-2W(Q)}
\frac{| \mathbf{Q} \cdot \mathbf{e}_{i,H} |^2}{\omega_i},
\label{eq:ins_intensity}
\end{equation}
where $\mathbf{Q}$ is the momentum transfer of the neutron, $\mathbf{e}_{i,H}$ denotes the displacement of hydrogen atoms in mode $i$, and $W(Q)$ is the Debye–Waller exponent. 
For an isotropic harmonic solid,
\begin{equation}
2W(Q) = Q^2 \langle u^2 \rangle,
\end{equation}
with $\langle u^2 \rangle$ the mean-square atomic displacement.
Equation~(\ref{eq:ins_intensity}) shows that, once Debye–Waller attenuation is accounted for, the measured peak intensity provides a direct experimental measure of hydrogen displacement weight. 
Because the diagonal frequency shift in Eq.~(\ref{eq:domega}) depends on the same projection of the eigenvector onto the mass perturbed subspace, the protiated INS intensity provides a quantitative predictor of the isotope-induced frequency renormalisation in spectrally isolated regions.

Beyond mere frequency shifts, changes in the eigenvectors arise from the off-diagonal matrix elements,
\begin{equation}
\delta\mathbf{q}_i
=
\sum_{j\neq i}
a_{ji}\,\mathbf{q}_j,
\label{eq:dq}
\end{equation}

\begin{equation}
a_{ji}
\equiv
\frac{
\mathbf{q}_j^{\mathrm T}\,\delta\mathbf{D}\,\mathbf{q}_i
}{
\omega_i^2-\omega_j^2
}.
\label{eq:aji}
\end{equation}
In spectrally congested regions, small frequency separations $\omega_i^2-\omega_j^2$ amplify mixing between neighbouring modes. 
If multiple nearby modes share substantial hydrogen displacement weight, the same mass perturbation that produces large diagonal shifts also enhances off-diagonal couplings, promoting collective redistribution of displacement character within a coupled manifold.

The redistribution observed in Fig.~\ref{snaps}B,E corresponds to moderate off-diagonal coupling within a relatively narrow manifold, whereas the qualitative change in Fig.~\ref{snaps}C,F reflects stronger mixing across a broader coupled subspace.

Experimentally, this behaviour can be diagnosed by direct comparison of the protiated and deuterated INS spectra. 
In congested regions, substantial redistribution of peak positions and relative intensities upon deuteration signals that neighbouring modes share significant hydrogen displacement weight and are therefore strongly coupled through Eq.~(\ref{eq:dq}). 
Conversely, minimal spectral redistribution within a dense cluster indicates that, despite small level spacings, the relevant eigenvectors remain weakly entangled under the applied mass perturbation.
\subsection{Adiabatic eigenvector continuation under mass deformation}
\label{sec:adiabatic}
To resolve ambiguities inherent in endpoint comparison, we introduce a continuous interpolation parameter $s\in[0,1]$ and evaluate the eigenproblem for $D(s)$ on a discrete grid $\{s_k\}$, with total grid points $N_{s}$.
\begin{equation}
\mathbf{M}(s)
=
(1-s)\,\mathbf{M}_{\mathrm H}
+
s\,\mathbf{M}_{\mathrm D},
\end{equation}
while the force-constant matrix $\boldsymbol{\Phi}$ remains fixed. The resulting mass-weighted dynamical matrix defines a one-parameter family of Hermitian operators,
\begin{equation}
\mathbf{D}(s)
=
\mathbf{M}(s)^{-1/2}
\boldsymbol{\Phi}
\mathbf{M}(s)^{-1/2}.
\label{eq:Ds}
\end{equation}
At each interpolation point $s_k$, we solve
\begin{equation}
\mathbf{D}(s_k)\,\mathbf{q}_i(s_k)
=
\omega_i^2(s_k)\,\mathbf{q}_i(s_k),
\end{equation}
with orthonormal mass-weighted eigenvectors satisfying
\(
\mathbf{q}_i^{\mathrm T}(s_k)\mathbf{q}_j(s_k)=\delta_{ij}.
\)
Because $\mathbf{D}(s)$ is Hermitian for all $s$, its eigenvalues are real and its eigenvectors form an orthonormal basis. In the absence of exact degeneracies, both eigenvalues and eigenvectors vary continuously with $s$.

In a strictly non-degenerate spectrum, frequency ordering would in principle be sufficient to define branch continuity. However, in dense vibrational manifolds it is not known a priori whether true crossings, avoided crossings, or higher-dimensional near-degeneracies occur. Moreover, even when eigenvalue ordering is preserved, substantial rotation of eigenvectors within near-degenerate subspaces can lead to redistribution of vibrational character that is not captured by frequency ordering alone. We therefore enforce continuity through a maximum-overlap assignment rule, which directly tracks eigenvector evolution and provides a diagnostic of subspace mixing.

For each eigenvector $\mathbf{q}_i(s_k)$, we compute the overlap with all eigenvectors at $s_{k+1}$,
\begin{equation}
O_{ij}(s_k)
=
\left|
\mathbf{q}_i^{\mathrm T}(s_k)\,
\mathbf{q}_j(s_{k+1})
\right|.
\end{equation}
We then define
\begin{equation}
O_i^{(1)}(s_k)
=
\max_j O_{ij}(s_k),
\end{equation}
as the largest overlap for mode $i$, and
\begin{equation}
O_i^{(2)}(s_k)
=
\max_{j \neq j^{(1)}} O_{ij}(s_k),
\end{equation}
as the second-largest overlap, where $j^{(1)}$ is the index that fulfills $O_i^{(1)}$.

The propagated eigenvector at $s_{k+1}$ is assigned to be the eigenvector associated with $O_i^{(1)}(s_k)$. This procedure implements a discrete version of adiabatic eigenvector continuation in the mass-weighted inner product.

For sufficiently small interpolation steps $\Delta s$, first-order perturbation theory gives
\begin{equation}
\mathbf{q}_i(s+\Delta s)
=
\mathbf{q}_i(s)
+
\sum_{j\neq i}
\frac{
\mathbf{q}_j^{\mathrm T}(s)\,
\frac{\partial \mathbf{D}}{\partial s}\,
\mathbf{q}_i(s)
}{
\omega_i^2(s)-\omega_j^2(s)
}
\,\Delta s\,\mathbf{q}_j(s)
+
\mathcal{O}(\Delta s^2).
\label{eq:perturb_local}
\end{equation}
Equation~(\ref{eq:perturb_local}) shows that eigenvector rotation is governed by off-diagonal matrix elements of $\partial\mathbf{D}/\partial s$, amplified when frequency separations become small. In the limit $\Delta s \rightarrow 0$, the maximum-overlap rule converges to continuous adiabatic evolution consistent with this perturbative form.

To quantify the stability of the propagation, we define the overlap gap
\begin{equation}
\Delta O_i(s_k)
=
O_i^{(1)}(s_k)
-
O_i^{(2)}(s_k).
\label{eq:gap}
\end{equation}

A large $\Delta O_i$ indicates unambiguous tracking of vibrational identity. When $\Delta O_i$ becomes small, either strong subspace mixing is occurring or the interpolation step is too coarse to resolve rapid eigenvector rotation.

When $|\omega_i^2(s)-\omega_j^2(s)|$ becomes comparable to the magnitude of off-diagonal couplings, eigenvectors undergo rapid but continuous rotation within a coupled subspace. If the interpolation spacing $\Delta s$ exceeds this intrinsic rotation scale, apparent crossings may arise from under-sampling. Refining the interpolation grid reduces this artefact and restores smooth eigenvector trajectories.

This framework provides a controlled and physically transparent method for tracking vibrational identity under continuous mass deformation and distinguishes genuine avoided-crossing behaviour from artefacts of coarse sampling.

\subsection{Continuity Diagnostic Results}
\begin{figure}[!h]
 \centering
 \includegraphics[width=0.95\linewidth]{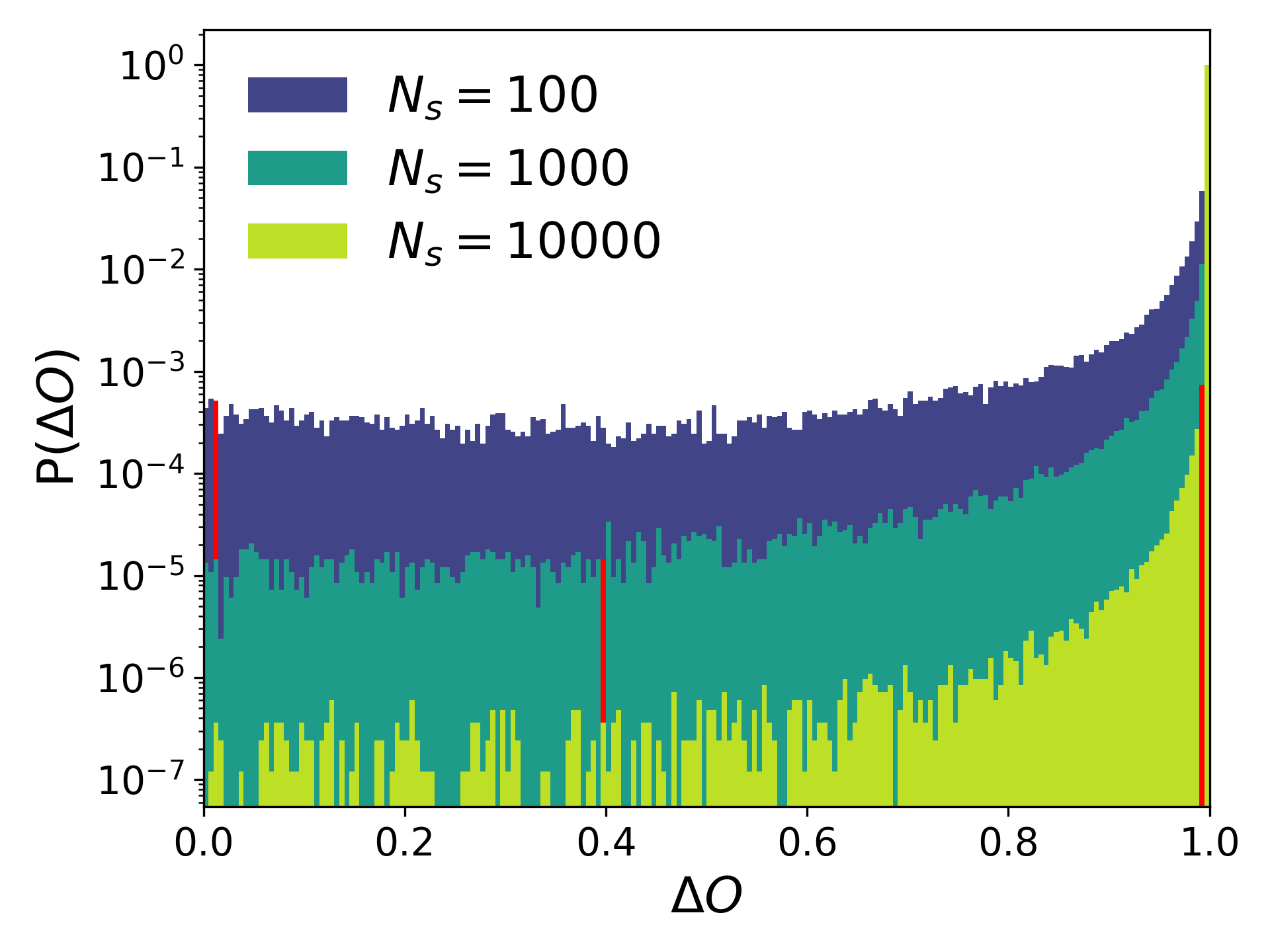}
 \caption{Normalized probability distributions $P(\Delta O)$ for interpolation grids 
$N_s = 100$, $1000$ and $10000$. 
Histograms are shown on a logarithmic scale 
to resolve the low-gap tail. The red bin marks the $\sim$0.1st percentile.}
 \label{probO}
\end{figure}

To quantify the robustness of the eigenvector-continuation procedure, calculations were performed using a range of values of $N_s$. Figure~\ref{probO} shows the variation of the normalized probability distributions of $\Delta O$ across three orders of magnitude of $N_{s}$. As the grid is refined, the long-tailed portion of the distribution progressively diminishes, indicating the suppression of low-gap (ambiguous) assignments. The 0.1st percentile of $P(\Delta O)$, shown as a red boundary in figure.\ref{probO}, shifts systematically toward higher values with increasing grid density; by $N_s = 10,000$, this percentile has already risen to $\Delta O \approx0.99$. This indicates that the vast majority of matches involve a large degree of separation from the next best candidate. 

\begin{figure}[!h]
 \centering
 \includegraphics[width=0.95\linewidth]{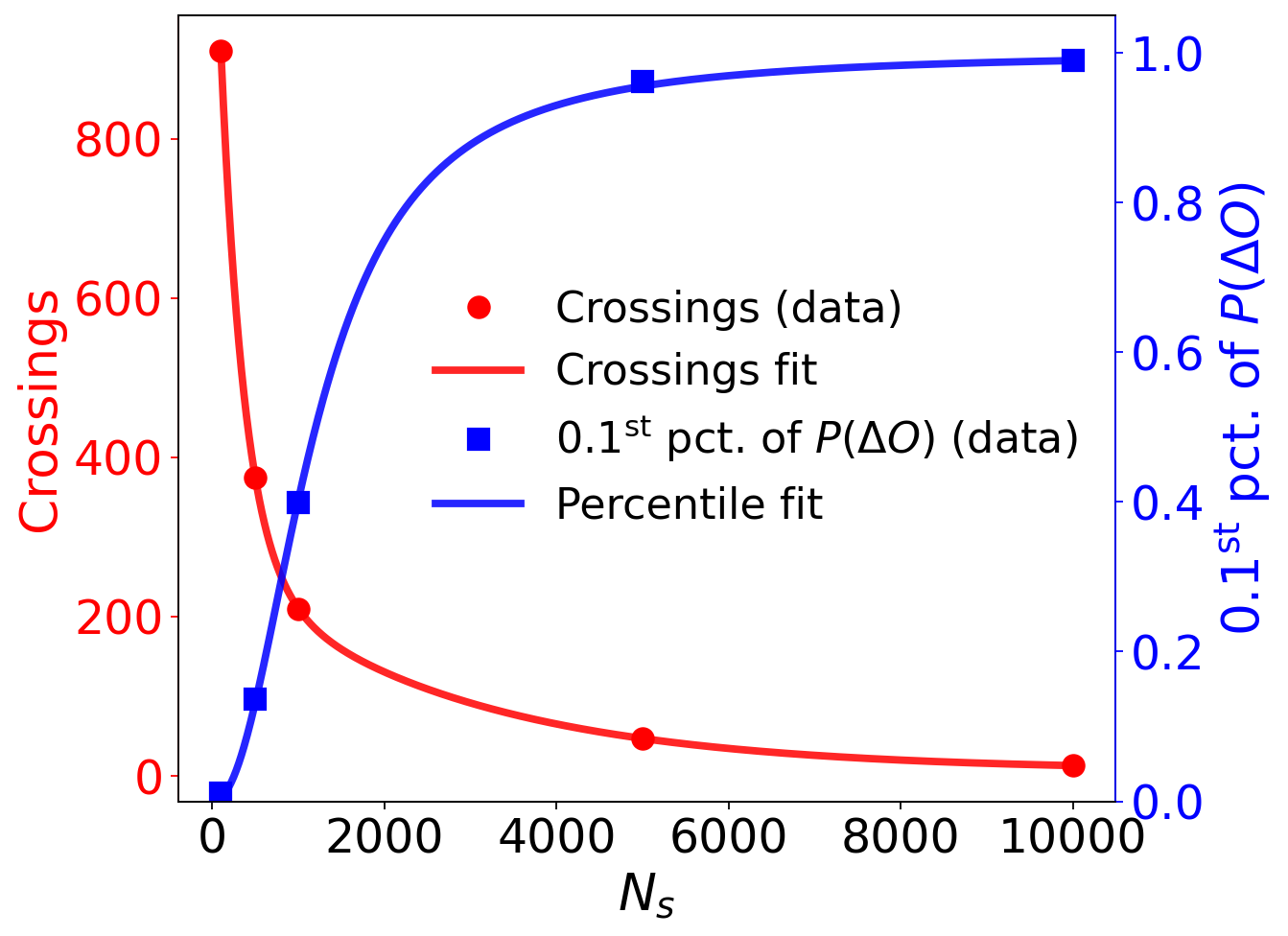}
 \caption{Grid-resolution dependence of spurious mode crossings and overlap robustness.
The number of detected crossings (red, left axis) decreases with increasing grid density $N_s$,
while the $0.1^{\mathrm{st}}$ percentile of $P(\Delta O)$ (blue, right axis)
increases toward unity, indicating improved continuity in mode tracking.
Solid lines show empirical fits to guide the eye.}
 \label{crossings}
\end{figure}

In parallel, we explicitly counted the number of apparent crossings identified under discrete propagation and plotted this quantity as a function of $N_s$ in Fig.~\ref{crossings}. The number of crossings decreases monotonically with increasing grid resolution, tending asymptotically toward zero, while the 0.1$^{st}$ percentile of $P(O\Delta)$ approaches unity. Taken together, these results demonstrate that the low-gap events and apparent crossings observed on coarse grids arise from under-sampling of rapid but continuous eigenvector rotation. Upon sufficiently fine interpolation, the evolution becomes unambiguous, consistent with avoided-crossing behaviour rather than true mode interchange.

\begin{figure}[!h]
 \centering
 \includegraphics[width=0.95\linewidth]{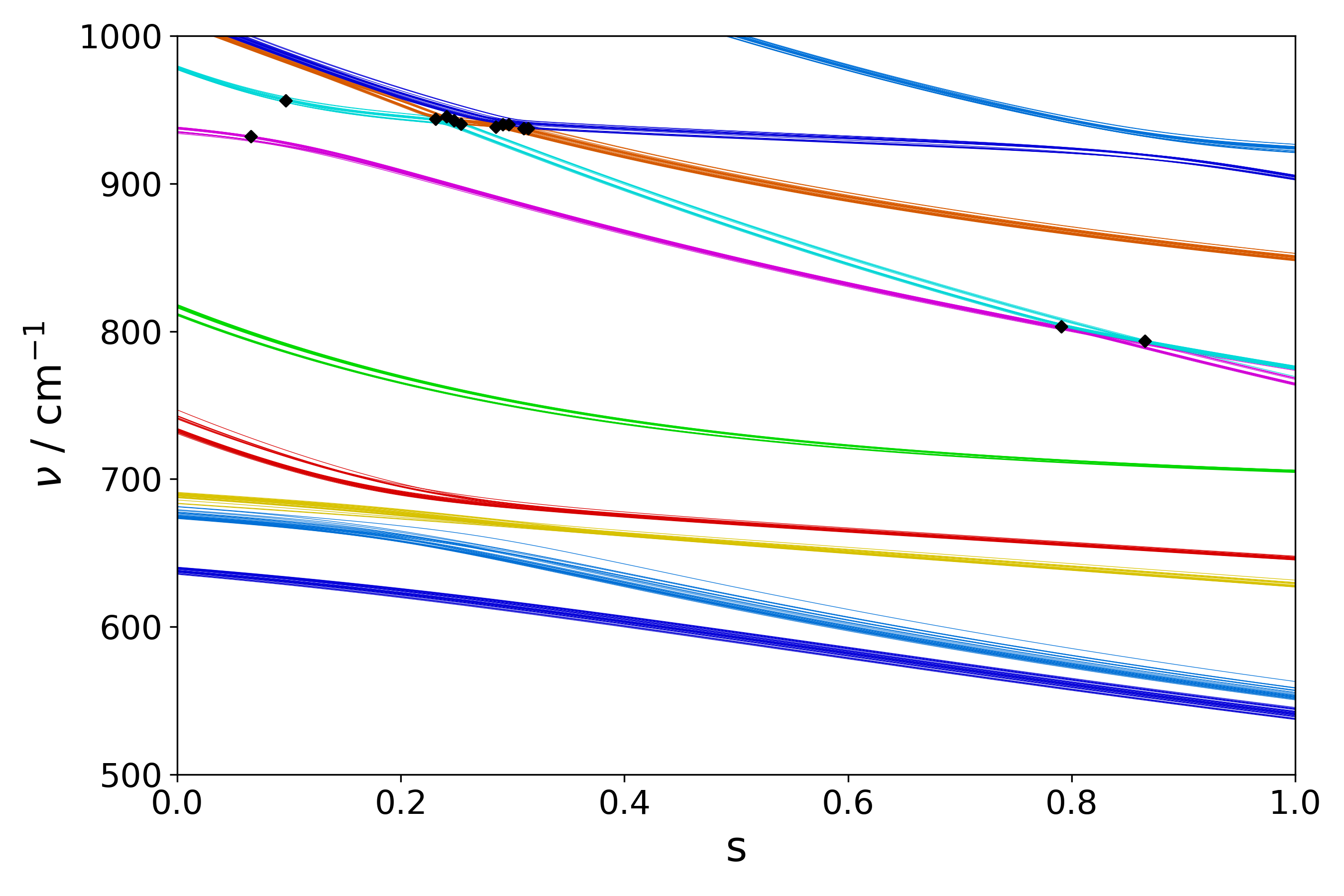}
 \caption{Continuously tracked phonon modes calculated on a dense interpolation grid of $N_s = 10{,}000$ points along the parameter coordinate $s$. Colours denote modes by their combined origin ($s=0$) and destination ($s=1$). Black diamonds indicate mode crossings.
}
 \label{spagh}
\end{figure}

Figure~\ref{spagh} shows the interpolated band diagram obtained with a refined grid of $N_s = 10{,}000$. At this resolution the majority of vibrational trajectories evolve smoothly across the full interpolation range. Distinct bands can be identified, connecting clusters of modes at $s=0$ to corresponding clusters at $s=1$. We assign colours to these bands based on the union of their initial and final clusters, where a cluster is defined as a set of modes separated by less than 10~cm$^{-1}$ in frequency. While many trajectories remain within their original cluster throughout the interpolation, others depart from their initial grouping (e.g.\ near $\sim 670~\mathrm{cm}^{-1}$) and/or merge into a different final cluster (e.g.\ near $\sim 550~\mathrm{cm}^{-1}$).

 With only 13 apparent crossings (black diamonds in figure\ref{crossings}), these residual cases are confined to the most heavily spectrally congested regions of the interpolation parameter $s$. In such dense regions the intrinsic rotation scale of the eigenvectors becomes small, so that coarse discretisation in $s$ fails to resolve the local curvature of the trajectory and can produce artificial branch interchange.

To further verify that these remaining crossings arise from insufficient resolution rather than genuine spectral discontinuities, we performed an additional targeted refinement in the restricted interval $s \in [0.2,\,0.35]$, where the majority of remaining residual crossings occur, using an additional $N_s = 6000$ grid points within this subdomain. This local refinement reduces the number of apparent crossings to 3, which represents a 3-fold decrease for an equivalent 4-fold increase in grid points within this region (9 crossings within this region for $N_{s}=10$,$000$, $s \in [0,\,1]$ ). This further confirms that the ambiguities are discretization artefacts and that the evolution is consistent with avoided-crossing behaviour.

Given the prohibitive computational cost of calculating individual matrix elements on a fine grid for all $s_{k}$, much of the remainder of the text will evaluate elements on a $N_{s}\leq10,000$ grid. Representative modes are chosen that do not undergo apparent ``crossings'', therefore ensuring mode continuity for the studied modes.

\subsection{Why a two-level ($2\times 2$) crossing picture is insufficient in dense manifolds}

The interpretation of apparent ``crossings'' is often motivated by the standard two-level reduction. When two eigenvalues approach one another, one projects the dynamical matrix onto the subspace spanned by the corresponding eigenvectors and studies an effective $2\times 2$ model. Writing $\lambda=\omega^2$, this reduction yields
\begin{equation}
D_{\mathrm{eff}}(s)=
\begin{pmatrix}
\lambda_i(s) & V_{ij}(s)\\
V_{ij}(s) & \lambda_j(s)
\end{pmatrix},
\end{equation}
with eigenvalues
\begin{equation}
\lambda_{\pm}(s)=\frac{\lambda_i+\lambda_j}{2}\pm 
\sqrt{\left(\frac{\lambda_i-\lambda_j}{2}\right)^2+V_{ij}(s)^2}.
\end{equation}
Within this model a true crossing requires simultaneously $\lambda_i=\lambda_j$ and $V_{ij}=0$ at some $s^\star$. Otherwise the minimum separation is finite and the interaction is an avoided crossing.

The numerical experiments presented above already demonstrate that the residual apparent crossings disappear upon sufficient refinement of the interpolation grid. However, the limitation of the $2\times 2$ picture in the present system is more fundamental. ZIF-8 exhibits spectrally congested manifolds/clusters, in which multiple eigenvalues remain mutually close over extended intervals of $s$. In such regions there typically exist several indices $k\neq i,j$ for which
\begin{equation}
|\lambda_i(s)-\lambda_k(s)| \lesssim |\lambda_i(s)-\lambda_j(s)|,
\end{equation}
so that projection onto $\mathrm{span}\{q_i,q_j\}$ is not dynamically closed. The eigenvector evolution is governed not by a single coupling $V_{ij}$ but by collective mixing within a near-degenerate subspace of dimension $m>2$. Many transport coefficients $a_{ji}(s)$ therefore
become simultaneously large for multiple $j$ as band spacings decrease, generating collective rapid rotation of $q_i(s)$ within a multi-dimensional cluster rather than pairwise level repulsion.

\subsection{Collective coupling, nonlocal mixing, and multi-band collisions}

We now examine the transport amplitudes $a_{ji}(s)$ directly for representative cases. These quantities provide a resolved view of eigenvector rotation under mass deformation and allow the mechanisms governing spectral flow to be visualised explicitly. As discussed above [Eq.~(\ref{eq:perturb_local})], the magnitude of $a_{ji}$ is controlled by two amplification channels: the spectral denominator $(\lambda_i-\lambda_j)^{-1}$ and the coupling matrix element $q_j^{\mathrm T}(\partial_s D)\,q_i$. The following examples illustrate how distinct spectral geometries activate these terms.

\paragraph*{Structural amplification: action at a distance.}

\begin{figure}[!h]
 \centering
 \includegraphics[width=0.95\linewidth]{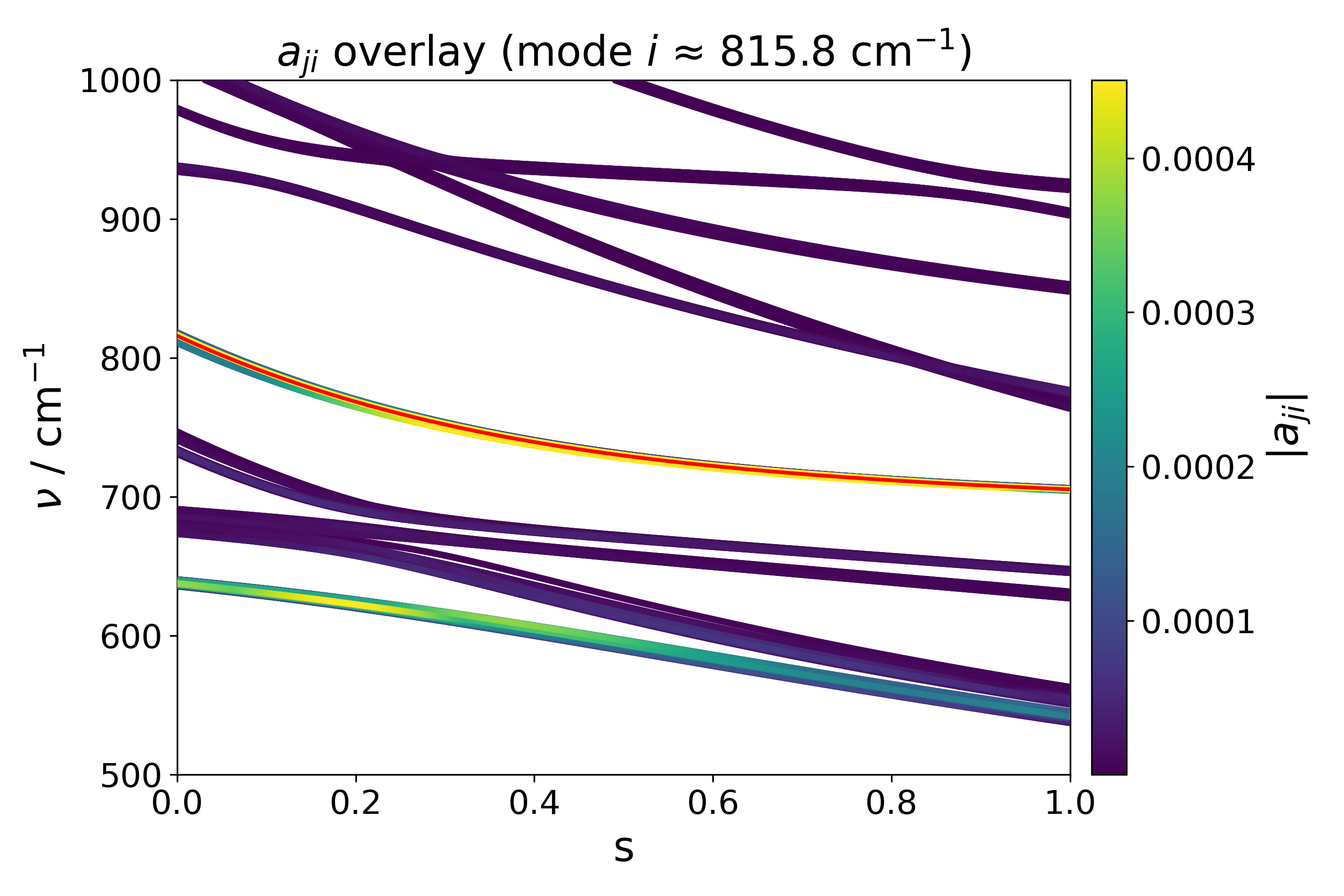}
 \caption{
$a_{ji}$ (color axis) for the tracked band ($\sim 816~\mathrm{cm^{-1}}$) as a function of $s$ and $\nu$.
A pronounced coupling is observed between this band and a non-adjacent band separated by three intermediate branches in frequency, demonstrating long-range mode interaction.
This behaviour illustrates ``action at a distance,'' whereby eigenvector character is transferred between spectrally separated bands despite the absence of nearest-neighbour proximity in frequency, facilitated by large mass-perturbation couplings between structurally similar displacement patterns.
}
 \label{actionatadistance}
\end{figure}

Figure~\ref{actionatadistance} shows $a_{ji}(s)$ for the tracked band $\nu_{s=0} \approx 816~\mathrm{cm^{-1}}$. A pronounced coupling is observed to a non-adjacent band separated by three intermediate bands in frequency. Because the frequency separation remains comparatively large, this behaviour is not driven by an anomalously small spectral denominator; instead, it reflects a large coupling matrix element. The two modes share displacement weight on out of phase, out of plane coupled C-H wag motions. These are isotope-sensitive coordinates, so $\partial_s D$ couples them strongly despite their spectral separation. This provides a clear example of ``action at a distance'', in which structural similarity rather than frequency proximity governs the transport. Visualisation of the mode in question shows that after following the mode to its isotopic end point ($s=1$) we see that the nature of the mode has completely changed. Its displacement pattern is now that of C-D in plane scissoring. This strong coupling to a much lower non-local mode has therefore imparted drastic qualitative change to the displacement pattern. 

\paragraph*{Three-band encounter.}

\begin{figure}[!h]
 \centering
 \includegraphics[width=0.95\linewidth]{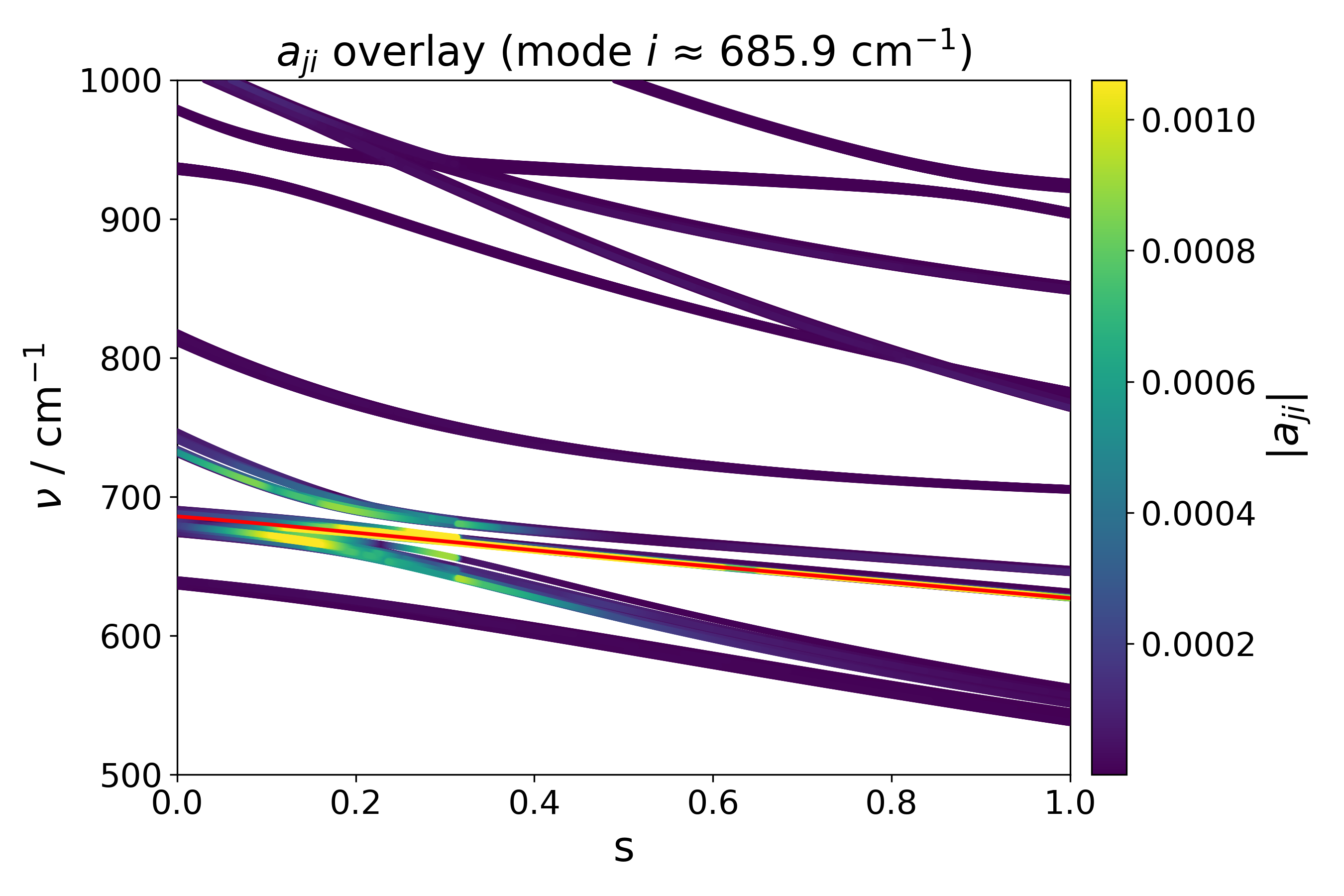}
 \caption{
Interaction amplitudes $a_{ji}(s)$ (color axis) for the selected mode $i\approx658.9 $~cm$^{-1}$(shown as a red line) within a three-band encounter region.
Coloured points denote the magnitude of $a_{ji}$ coupling to other modes.
Enhanced amplitudes emerge as the three branches approach in frequency, indicating gradual coordinated eigenvector rotation.
}
 \label{intermedexample}
\end{figure}

Beyond this two-band nonlocal interaction, we observe intermediate situations in which three trajectories approach within a moderately narrow spectral window (Fig.~\ref{intermedexample}). As their squared frequencies evolve at unequal rates under mass deformation, spectral separations are reduced and multiple denominator terms become enhanced simultaneously. The intermediate branch responds to both neighbours, producing coordinated redistribution of vibrational character within a locally coupled three-dimensional subspace. This regime already exceeds a simple pairwise avoided-crossing picture but remains confined to a small cluster. 

\paragraph*{Collective rotation in a congested manifold.}

\begin{figure}[!h]
 \centering
 \includegraphics[width=0.95\linewidth]{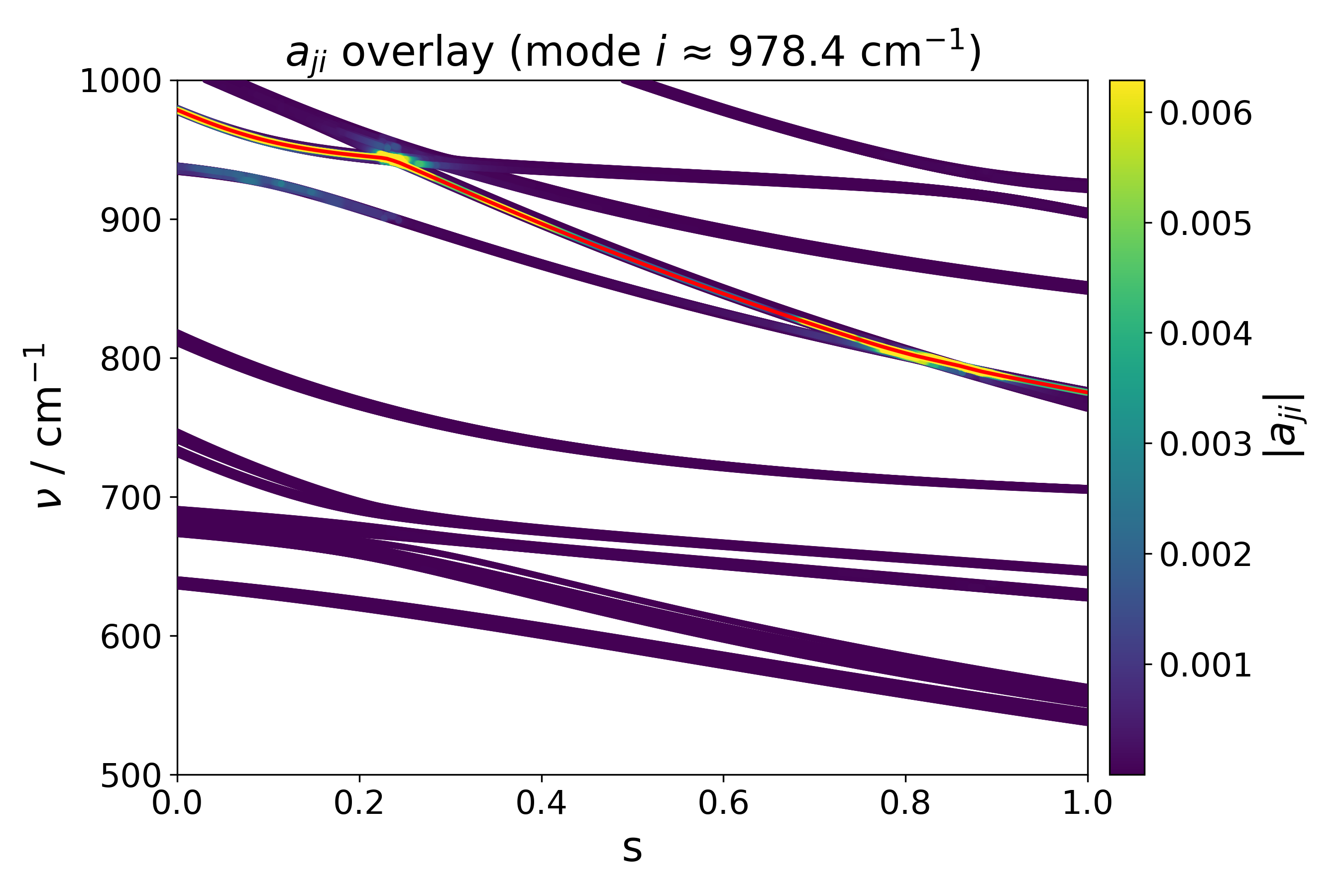}
 \caption{
Transport amplitudes $a_{ji}(s)$ (color axis) in a densely congested manifold, illustrating highly localised collective multi-mode mixing.
As multiple branches approach within a narrow spectral window, the transport amplitude becomes distributed across a cluster of nearby modes, consistent with rapid rotation within a higher-dimensional near-degenerate subspace rather than isolated pairwise interactions.
}
 \label{congestaji}
\end{figure}

In the most densely populated regions, several eigenvalues remain mutually close over extended intervals of $s$. In this regime multiple denominator terms in Eq.~(\ref{eq:perturb_local}) become simultaneously large, and the transport amplitudes become distributed across a cluster of nearby modes. Figure~\ref{congestaji} shows a representative example in which mixing with the above band is initially weak in relative terms. This mode descends through a first order frequency shift, and upon close frequency contact, the three bands are forced to undergo much more aggressive eigenvector rotations to prevent a crossing. This fine scale rearrangement indicates an initial poorly projected mass perturbation overlap between the approaching modes, where the rapid eigenvector rotation is driven instead by the $\omega_i^2(s)-\omega_j^2(s)$ denominator.

Taken together, these examples show that vibrational reorganisation under isotopic mass deformation is governed by a delicate interplay between first order frequency shifts initiated by strong displacement patterns in the mass substituted site, spectral amplification (small denominators) and structural amplification (large coupling matrix elements). The resulting dynamics are intrinsically multi-level and cannot be reduced to isolated two-state interactions. Vibrational identity in dense framework materials is therefore defined by continuous transport of invariant subspaces rather than by pairwise frequency matching.

\section{Conclusions and predictive outlook}
We have shown that isotope substitution defines a controlled mass deformation of the dynamical matrix and therefore provides a direct experimental probe of vibrational transport geometry. Our conclusions show crucially that the comparison of neutron spectra of protiated and deuterated samples does more than reveal frequency shifts. Changes in peak distribution and intensity encode information about qualitative redistribution of eigenvector character.

In spectrally sparse regions of a protiated INS spectrum, the qualitative and quantitative consequences of deuteration can be anticipated directly from the experimental data. 
Equation~(\ref{eq:domega}) shows that the first-order frequency shift under mass substitution is governed by the projection of the eigenvector onto the mass perturbed subspace. For modes dominated by hydrogen motion, the mass-weighted displacement amplitude on hydrogen sites is large, and the resulting frequency shift upon deuteration is correspondingly substantial.
For INS, the measured intensity reflects hydrogen displacement amplitudes weighted by the scattering cross section and modulated by the Debye–Waller factor. Once the Debye-waller contribution is accounted for, the relative intensity of isolated peaks provides a direct experimental measure of hydrogen participation and hence a quantitative predictor of the isotope-induced frequency shift.

A complementary predictive insight arises in spectrally congested regions. 
Here the isotope response is governed jointly by the diagonal frequency renormalisation [Eq.~(\ref{eq:domega})] and the eigenvector transport amplitudes [Eq.~(\ref{eq:dq})]. 
When multiple neighbouring modes possess significant displacement weight on the perturbed mass subspace, small level spacings amplify off-diagonal couplings and promote collective subspace rotation.
Experimentally, comparison of protiated and deuterated spectra in such congested regions provides a diagnostic of this mixing. 
Substantial redistribution of peak positions and relative intensities upon deuteration signals strong shared displacement character among neighbouring modes and hence susceptibility to qualitative eigenvector reorganisation. Conversely, minimal redistribution within a dense spectral cluster indicates that, despite spectral crowding, the relevant displacement patterns remain weakly coupled under the applied mass perturbation.
These concepts sugggest thereofre that deuteration can in principle be used as a means to shift undesirable modes in frequency, or to alter their qualitative displacement character completely. In particular we could imagine qualitatively altering ``killer'' modes in organic semiconductors, or tuning gate-opening mechanisms in flexible metal–organic frameworks. In fact, even greater control may be achieved through selective deuteration of specific molecular groups. This targeted approach creates the flexibility to apply changes to given displacement patterns, while leaving other functionally beneficial ones unchanged. With the information provided by INS, combined with these controlled isotope substitutions, we arrive at a strategy for deliberate phonon steering and controlled redistribution of vibrational displacement character.

\section{Acknowledgments}
We would like to thank Dr. Srinivasu Kancharlapalli of the Chemsitry Divison, Bhabha Atomic Research Centre, Mumbai for their assitance with the DFT calculations that formed the basis of the force-constant matrix. Explicit credit also goes to P.S. for synthesizing the INS samples, which strongly validate this work. This work was supported by the UKAEA / EPSRC Fusion Grant 2022/27 (P/W006839/1), UK Research and Innovation.

CRediT author statement: 
J.A.: Formal analysis, Investigation, Methodology, Conceptualization, Software, Data curation, Visualization, Validation, Writing - original draft. 
H. C.: Conceptualization, Investigation, Methodology, Data curation, Visualization, Writing - review \& editing.
P. S.: Conceptualization, Funding acquisition, Investigation, Resources, Project administration, Data curation, Writing - review \& editing.
M. E. P.: Conceptualization, Funding acquisition, Investigation, Resources, Project administration, Writing - review \& editing.

\appendix

\section{Derivation of main equations}
\label{sec:mainderive}
\subsection{Harmonic lattice dynamics and physical displacement fields}

We consider a system of $N$ atoms with equilibrium positions $\{\mathbf{R}_a\}$.
Let
\[
\mathbf{u}(t) = \{u_{a\alpha}(t)\}
\]
denote the Cartesian displacement vector, where $u_{a\alpha}$ is the displacement of
atom $a$ along Cartesian direction $\alpha \in \{x,y,z\}$.

Within the harmonic approximation, the potential energy expanded about equilibrium is
\begin{equation}
V(\mathbf{u})
=
V_0
+
\frac{1}{2}
\sum_{a\alpha,b\beta}
\Phi_{a\alpha,b\beta}\,
u_{a\alpha}\,u_{b\beta},
\end{equation}
where $\boldsymbol{\Phi}$ is the force-constant matrix with elements
\begin{equation}
\Phi_{a\alpha,b\beta}
=
\frac{\partial^2 V}{\partial u_{a\alpha}\,\partial u_{b\beta}}.
\end{equation}

The equations of motion are
\begin{equation}
m_a\,\ddot{u}_{a\alpha}(t)
=
-
\sum_{b\beta}
\Phi_{a\alpha,b\beta}\,u_{b\beta}(t),
\end{equation}
which may be written compactly as
\begin{equation}
\mathbf{M}\,\ddot{\mathbf{u}}(t)
=
-
\boldsymbol{\Phi}\,\mathbf{u}(t),
\end{equation}
where $\mathbf{M}$ is the diagonal mass matrix.

\subsection{Normal modes and mass-weighted coordinates}

We seek harmonic solutions of the form
\begin{equation}
\mathbf{u}(t) = \mathbf{u}_i\,e^{i\omega_i t},
\end{equation}
where $\mathbf{u}_i$ is a time-independent displacement pattern and $\omega_i$ is the
angular frequency of mode $i$.

Substitution yields the generalised eigenvalue problem
\begin{equation}
\boldsymbol{\Phi}\,\mathbf{u}_i
=
\omega_i^2\,\mathbf{M}\,\mathbf{u}_i.
\label{eq:SI_gen_eig}
\end{equation}

Introducing mass-weighted displacement vectors
\begin{equation}
\mathbf{q}_i \equiv \mathbf{M}^{1/2}\mathbf{u}_i,
\end{equation}
transforms the problem into the standard Hermitian eigenvalue equation
\begin{equation}
\mathbf{D}\,\mathbf{q}_i
=
\omega_i^2\,\mathbf{q}_i,
\label{eq:SI_massweighted_eig}
\end{equation}
where the mass-weighted dynamical matrix is
\begin{equation}
\mathbf{D}
=
\mathbf{M}^{-1/2}\boldsymbol{\Phi}\mathbf{M}^{-1/2}.
\label{eq:SI_D_def}
\end{equation}

The eigenvectors satisfy the orthonormality condition
\begin{equation}
\mathbf{q}_i^{\mathrm T}\mathbf{q}_j = \delta_{ij}.
\end{equation}

All modes considered in this work correspond to $\Gamma$-point normal modes and may be
chosen real.

\subsection{Isotope substitution as a mass perturbation}

Isotope substitution modifies the atomic masses while leaving the force-constant matrix
unchanged to first order,
\begin{equation}
\mathbf{M} \rightarrow \mathbf{M} + \delta\mathbf{M},
\qquad
\boldsymbol{\Phi} \ \text{fixed}.
\end{equation}
Here $\delta\mathbf{M}$ is diagonal and nonzero only on substituted degrees of freedom.

This induces a perturbation of the dynamical matrix
\begin{equation}
\mathbf{D} \rightarrow \mathbf{D} + \delta\mathbf{D},
\end{equation}
through the mass-weighting in Eq.~\eqref{eq:SI_D_def}.

\subsection{Derivation of the mass-induced perturbation $\delta\mathbf{D}$}

Starting from Eq.~\eqref{eq:SI_D_def}, we write
\begin{equation}
\mathbf{D}(\mathbf{M}) = \mathbf{M}^{-1/2}\,\boldsymbol{\Phi}\,\mathbf{M}^{-1/2}.
\end{equation}
Under $\mathbf{M}\to \mathbf{M}+\delta\mathbf{M}$, the first-order change is
\begin{equation}
\delta\mathbf{D}
=
\delta(\mathbf{M}^{-1/2})\,\boldsymbol{\Phi}\,\mathbf{M}^{-1/2}
+
\mathbf{M}^{-1/2}\,\boldsymbol{\Phi}\,\delta(\mathbf{M}^{-1/2}),
\label{eq:SI_deltaD_chain}
\end{equation}
where we have used $\delta\boldsymbol{\Phi}=0$.

Define the dimensionless diagonal matrix
\begin{equation}
\mathbf{G} \equiv \mathbf{M}^{-1/2}\,\delta\mathbf{M}\,\mathbf{M}^{-1/2}.
\label{eq:SI_G_def}
\end{equation}
For a diagonal mass matrix, the first-order variation of $\mathbf{M}^{-1/2}$ is
\begin{equation}
\delta(\mathbf{M}^{-1/2})
=
-\frac{1}{2}\,\mathbf{M}^{-1/2}\,\mathbf{G}.
\end{equation}
Substituting into Eq.~\eqref{eq:SI_deltaD_chain} gives
\begin{equation}
\boxed{
\delta\mathbf{D}
=
-\frac{1}{2}\left(
\mathbf{G}\mathbf{D}
+
\mathbf{D}\mathbf{G}
\right).
}
\label{eq:SI_deltaD_final}
\end{equation}

\subsection{First-order perturbation theory for frequencies and eigenvectors}

At a fixed interpolation point, the unperturbed eigenproblem is
\begin{equation}
\mathbf{D}\,\mathbf{q}_i = \omega_i^2\,\mathbf{q}_i,
\qquad
\mathbf{q}_i^{\mathrm T}\mathbf{q}_i = 1.
\label{eq:SI_unperturbed}
\end{equation}

After a small mass perturbation, we write
\begin{equation}
\mathbf{D}\to \mathbf{D}+\delta\mathbf{D},\qquad
\omega_i^2\to \omega_i^2+\delta\omega_i^2,\qquad
\mathbf{q}_i\to \mathbf{q}_i+\delta\mathbf{q}_i.
\end{equation}

Substituting into Eq.~\eqref{eq:SI_unperturbed} and retaining only first-order terms gives
\begin{equation}
\mathbf{D}\,\delta\mathbf{q}_i + \delta\mathbf{D}\,\mathbf{q}_i
=
\omega_i^2\,\delta\mathbf{q}_i + \delta\omega_i^2\,\mathbf{q}_i.
\label{eq:SI_first_order_eq}
\end{equation}

\paragraph{First-order shift in $\omega_i^2$.}
Left-multiplying Eq.~\eqref{eq:SI_first_order_eq} by $\mathbf{q}_i^{\mathrm T}$ yields
\begin{equation}
\boxed{
\delta\omega_i^2
=
\mathbf{q}_i^{\mathrm T}\,\delta\mathbf{D}\,\mathbf{q}_i.
}
\end{equation}

\paragraph{First-order change in $\mathbf{q}_i$.}
Expanding
\begin{equation}
\delta\mathbf{q}_i = \sum_j c_{ji}\,\mathbf{q}_j,
\end{equation}
and enforcing $\mathbf{q}_i^{\mathrm T}\delta\mathbf{q}_i=0$ gives
\begin{equation}
\delta\mathbf{q}_i = \sum_{j\neq i} c_{ji}\,\mathbf{q}_j.
\end{equation}
Projecting Eq.~\eqref{eq:SI_first_order_eq} onto $\mathbf{q}_j$ with $j\neq i$ yields
\begin{equation}
(\omega_j^2-\omega_i^2)c_{ji}
=
\mathbf{q}_j^{\mathrm T}\,\delta\mathbf{D}\,\mathbf{q}_i,
\end{equation}
and therefore
\begin{equation}
\boxed{
\delta\mathbf{q}_i
=
\sum_{j\neq i}
\frac{
\mathbf{q}_j^{\mathrm T}\,\delta\mathbf{D}\,\mathbf{q}_i
}{
\omega_i^2-\omega_j^2
}
\,\mathbf{q}_j.
}
\label{eq:SI_dq_final}
\end{equation}

\subsection{Mode-resolved coupling and mixing amplitudes}

The physically fundamental scalar coupling induced by the mass perturbation between
modes $j$ and $i$ is
\begin{equation}
\boxed{
N_{ji}
=
\mathbf{q}_j^{\mathrm T}\,
\delta\mathbf{D}\,
\mathbf{q}_i.
}
\end{equation}

The corresponding first-order mixing amplitude is defined as
\begin{equation}
\boxed{
a_{ji}
=
\frac{N_{ji}}{\omega_i^2-\omega_j^2}.
}
\end{equation}

With this definition,
\begin{equation}
\delta\mathbf{q}_i
=
\sum_{j\neq i}
a_{ji}\,\mathbf{q}_j.
\end{equation}
\section{Convergence fit functions and parameters}

\begin{equation}
\mathrm{Crossings}(N_s)
= A_1 e^{-k_1 N_s}
+ A_2 e^{-k_2 N_s}
+ C.
\end{equation}

\begin{align}
A_1 &= 9.4 \times 10^{2}, &
k_1 &= 3.6 \times 10^{-3}, \\
A_2 &= 2.6 \times 10^{2}, &
k_2 &= 3.7 \times 10^{-4}, \\
C &= 6.8.
\end{align}

\begin{equation}
P(N_s)
= \frac{1}{1+\left(\dfrac{N_0}{N_s}\right)^{\beta}}.
\end{equation}

\begin{align}
N_0 &= 1.2 \times 10^{3}, \\
\beta &= 2.1.
\end{align}

\bibliographystyle{apsrev4-2}
\bibliography{references}

\end{document}